\documentclass[twocolumn,superscriptaddress,,preprintnumbers,amsmath,amssymb,prl]{revtex4}\usepackage{graphicx,bm}
\usepackage[dvips]{color}
\begin{document}
\title{Superconducting properties of the In substituted Topological Crystalline Insulator, SnTe}
\author{G. Balakrishnan}
\affiliation{Department of Physics, University of Warwick, Coventry, CV4 7AL, UK}

\author{L. Bawden}
\affiliation{Department of Physics, University of Warwick, Coventry, CV4 7AL, UK}

\author{S. Cavendish}
\affiliation{Department of Physics, University of Warwick, Coventry, CV4 7AL, UK}

\author{M. R. Lees}
\affiliation{Department of Physics, University of Warwick, Coventry, CV4 7AL, UK}

\date{\today}

\begin{abstract}
We report detailed investigations of the properties of a superconductor obtained by substituting In at the Sn site in the topological crystalline insulator (TCI), SnTe. Transport, magnetization and heat capacity measurements have been performed on crystals of Sn$_{0.6}$In$_{0.4}$Te,  which is shown to be a bulk superconductor with $T_c^{\rm{onset}}$ at $\sim4.70(5)$~K and $T_c^{\rm{zero}}$ at $\sim3.50(5)$~K. The upper and lower critical fields are estimated to be $\mu_0H_{c2}(0)=1.42(3)$~T and $\mu_0H_{c1}(0)=0.90(3)$~mT respectively, while $\kappa=56.4(8)$ indicates this material is a strongly type II superconductor. 
\end{abstract}
\maketitle

Topological insulators (TIs) have been extensively investigated in the recent past~\cite{Hasan, Qi} leading to the rapid discovery of three dimensional TIs (such as Bi$_2$Se$_3$, Bi$_2$Te$_3$), two dimensional TIs (such as HgTe, Hg-Cd-Te) and TIs that can be made superconducting, such as Cu$_x$Bi$_2$Se$_3$. More recently, a new class of materials called topological crystalline insulators (TCI) has been proposed theoretically by Fu ~\cite{Fu} and subsequently, SnTe has been discovered to exhibit all the required characteristics of this state~\cite{Tanaka, Hsieh}.
In a topological insulator the surface exhibits a gapless metallic state that is protected by time reversal symmetry. The surface states when examined by angle-resolved photo emission spectroscopy (ARPES) studies, reveal the presence of an odd number of Dirac cones. In the SnTe TCI, the topologically protected metallic surface states are protected by the mirror symmetry of the crystal, which replaces the role played by time reversal symmetry in protecting the surface states in a TI. The presence of band inversion in SnTe is also responsible for the observed surface states and differentiates this material from the analogous PbTe, which is not a TCI~\cite{Tanaka}. In the Pb$_{1-x}$Sn$_{x}$Se, the system is a TCI for $x=0.23$ and the material undergoes a transition from a trivial insulator to a TCI as a function of temperature~\cite{Dziawa}.

The study of TIs as superconductors has been less straightforward. Both surface as well bulk studies on the copper intercalated Bi$_2$Se$_3$, Cu$_x$Bi$_2$Se$_3$~\cite{Hor}, have produced results that are far from being consistent due to the intrinsic inhomogeneity of the samples. The half Heusler alloys YPtBi, LaPtBi  and LuPtBi are classified as TIs and are also known to be superconducting~\cite{Butch, Goll, Tafti}. Recently, these materials have also been identified as possible candidates for 3D topological superconductors based on the band inversion that they exhibit~\cite{Butch, Tafti}.

SnTe crystallizes in the rock salt structure and is classified a TCI because it satisfies the conditions required for the mirror symmetry. It is a narrow band semiconductor and in the as-grown state, it usually forms with a number of Sn vacancies. SnTe can be made superconducting when doped with about $10^{20}$~cm$^{-3}$ Sn vacancies ($T_c < 0.3$~K). It has been shown previously that SnTe when doped with small levels of In at the Sn site exhibits superconductivity~\cite{Erickson}. Previous work by Erickson \textit{et al}.~\cite{Erickson2} has shown that a superconductor with a maximum $T_c$ of 2~K is obtained for In substitution levels of around 6\% in Sn$_{0.988-x}$In$_{x}$Te, and detailed bulk property measurements have been reported on In substituted SnTe crystals, for low levels of substitution. Point contact spectroscopy, ARPES and bulk property measurements have been reported on Sn$_{1-x}$In$_{x}$Te for levels of In substitution giving superconducting transition temperatures between 1 and 2~K.~\cite{Sasaki, Sato}. More recently, it has been demonstrated that higher levels of In substitution ($x\sim0.4$) in SnTe results in a superconductor with a much higher transition temperature of $\sim4.5$~K ~\cite{APSSNTE}. These superconductors are particularly interesting as they emerge from a parent TCI material. The investigation of both the bulk and surface characteristics of these materials is essential in order to understand the electronic properties and their implications to the emergence of the superconducting state. 

In this paper, we report the synthesis and properties of  Sn$_{1-x}$In$_{x}$Te, for $x=0$, and 0.4 where $x=0.4$ sample is the superconductor and the $x=0$ sample is the parent non superconducting TCI for comparison. The In substitution level chosen for study here, appears to give close to the optimum $T_c$ for this system. The investigation of the superconducting properties  of the Sn$_{0.6}$In$_{0.4}$Te superconductor through resistivity, dc magnetization, ac susceptibility and heat capacity measurements are presented. The results point to the existence of bulk superconducting states in this important class of materials. 

Crystals of Sn$_{1-x}$In$_{x}$Te  for $x=0$ and 0.4, were grown by the modified Bridgman method adopting a similar procedure to that described by Tanaka \textit{et al}.~\cite{Tanaka} for SnTe. Stoichiometric ratios of the starting materials, 99.99\% Sn,(shots) In (shots) and Te (powder), were taken in evacuated and sealed quartz ampoules. The quartz tubes were heated to around $900~^{\circ}$C and slow cooled ($2~^{\circ}$C/h) to $770~^{\circ}$C,  followed by a fast cool to room temperature.   
Powder x-ray diffraction on powdered portions of the as grown boules were carried out using a Panalytical X' Pert Pro system with a monochromatic Cu$K_{\alpha1}$ radiation.

Rectangular bars ($\sim4$ to 6~mm in length and $\sim3\times1$~mm$^2$ in cross section) were cut for resistivity measurements. Resistivity $\rho$, was measured as a function of temperature (300 to 1.8~K) and applied magnetic field (0 to 2~T) using a standard 4 probe method in a Quantum Design Physical Property Measurement System (PPMS). Measurements of the sample heat capacity $C$, were carried out using a relaxation method in a Quantum Design PPMS. Measurements of ac susceptibility $\chi_{\rm{ac}}$, and dc magnetization $M$, were made as a function of temperature $T$, and applied magnetic field $H$, using a Quantum Design Magnetic Property Measurement System (MPMS).  
 
The observed x-ray powder diffraction pattern for the Sn$_{0.6}$In$_{0.4}$Te is consistent with the pattern for the SnTe ($x=0$) parent phase. Both materials have a cubic $Fm\bar{3}m$ structure. The lattice parameter calculated from a fit to the data is $a = 0.6280(1)$~nm  for Sn$_{0.6}$In$_{0.4}$Te, while the parent compound SnTe has a lattice parameter of $a = 0.6321(1)$~nm. Our results agree well with the published data for both the $x=0$ and the $x=0.4$~\cite{Zhigarev}. 

\begin{figure}[tb]
\begin{center}\includegraphics[width=0.9\columnwidth]{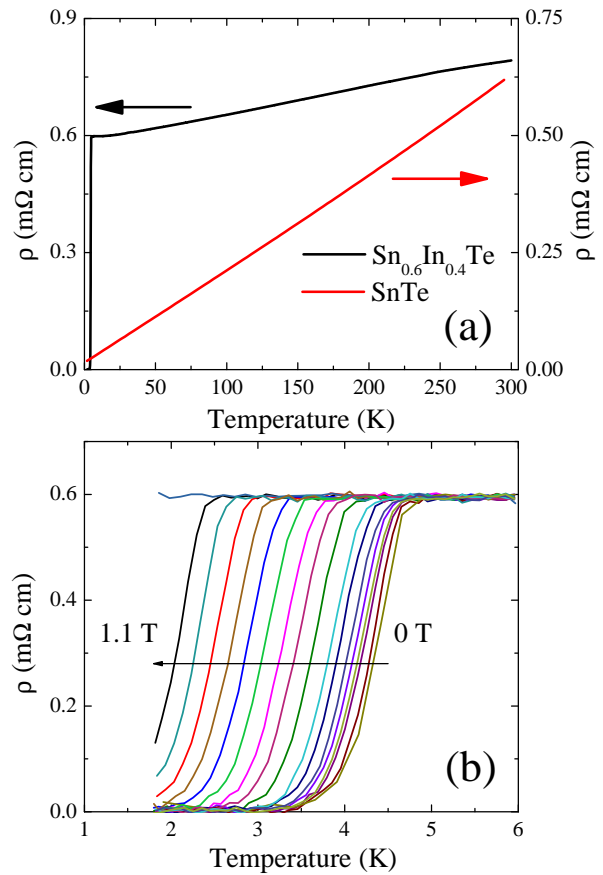}\caption{\label{Fig1} (Color online). (a) Temperature dependence of the resistivity of Sn$_{0.6}$In$_{0.4}$Te and SnTe in zero field. Sn$_{0.6}$In$_{0.4}$Te shows a superconducting transition with a $T_c^{\rm{onset}}$ at $\sim4.70(5)$~K and zero resistance, $T_c^{\rm{zero}}$ at ~3.50(5)~K.  (b) The resistivity of superconducting Sn$_{0.6}$In$_{0.4}$Te at temperatures around $T_c$ in increasing applied magnetic fields. The onset of the transition shifts to lower temperatures with increasing field without any broadening of the transition.}
\end{center}
\end{figure}

\begin{figure}[tb]
\begin{center}
\includegraphics[width=0.9\columnwidth]{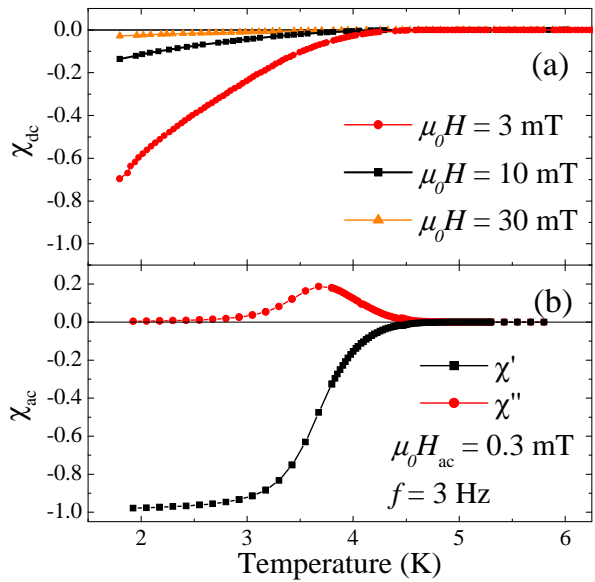}
\caption{\label{Fig2} (Color online). (a) Temperature dependence of the dc magnetic susceptibility of Sn$_{0.6}$In$_{0.4}$Te measured by zero-field-cooled warming (ZFCW). (b) Temperature dependence of the ac susceptibility of Sn$_{0.6}$In$_{0.4}$Te showing $\chi'_{\rm{ac}}$ and $\chi''_{\rm{ac}}$ through the superconducting transition. A demagnetization factor has been applied to account for the plate-like shape the sample~\cite{Aharoni}.}
\end{center}
\end{figure}

The temperature dependence of the resistivities of the $x=0.4$ superconducting sample and the non-superconducting SnTe ($x=0$) are shown in Fig.~\ref{Fig1}a.  In zero field, the superconducting transition temperature of Sn$_{0.6}$In$_{0.4}$Te, $T_c^{\rm{onset}}$ is $4.70(5)$~K. The width of the transition, $\Delta T_c$, is over 1~K with zero resistance indicating the presence of continuous superconducting pathway observed at $T_c^{\rm{zero}}=3.50(5)$~K. The $x=0$ non-superconducting sample shows an almost linear dependence of resistivity with $T$ over the entire range of the measurement. 

The resistivity of the $x=0.4$ sample at temperatures around $T_c$ in different applied fields is shown in Fig.~1b. While there is a gradual shift in $T_c^{\rm{onset}}$ with applied field, there is no broadening of the superconducting transition.

\begin{figure}[tb]
\begin{center}
\includegraphics[width=0.9\columnwidth]{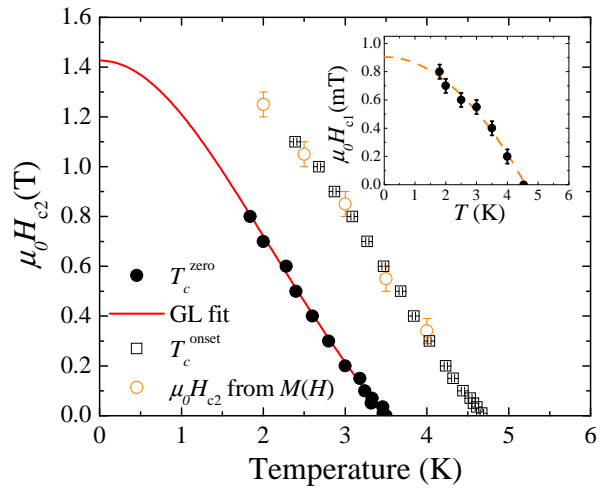}
\caption{\label{Fig3} (Color online). Temperature dependence of the upper critical field of Sn$_{0.6}$In$_{0.4}$Te. The solid symbols were determined from $T_c^{\rm{zero}}$ in the $\rho(T)$ data and the solid line is a fit to the data using a GL model (see text) giving a $\mu_0H_{c2}(0) = 1.42(3)$~T. The open symbols were determined from $T_c^{\rm{onset}}$ in the $\rho(T)$ data and $H_{\rm{c2}}$ values from $M(H)$ loops collected at fixed $T$. The inset shows $T$ dependence of the lower critical field $H_{\rm{c1}}(T)$. The dashed line is a fit assuming a parabolic $T$ dependence giving $H_{\rm{c1}}(0)=0.90(3)$~mT.}
\end{center}
\end{figure}

A superconducting transition in Sn$_{0.6}$In$_{0.4}$Te was also observed in dc magnetic susceptibility ($\chi_{\rm{dc}}=M/H$) measurements, which were carried out as a function of temperature in applied fields from 3~mT to 0.8~T. Fig.~\ref{Fig2}a shows the zero-field-cooled warming data for three applied magnetic fields. The magnetization observed in the superconducting state does not saturate at the lowest temperature measured. The lower critical field $H_{c1}$ is difficult to determine.  We estimate $H_{c1}(T)$ of this superconductor from the first deviation from linearity in the low-field regions in our $M(H)$ scans (not shown). Assuming a simple parabolic $T$ dependence for $H_{c1}(T) = H_{c1}(0)\left(1-t^2\right)$ where $t = T/T_c$ we estimate $H_{c1}(0)$ to be of the order of $0.90(3)$~mT (see inset of Fig.~\ref{Fig3}). The temperature dependence of the ac susceptibility $\chi_{\rm{ac}}$, measured on the same Sn$_{0.6}$In$_{0.4}$Te sample is shown in Fig~\ref{Fig2}b. For both the ac and dc measurements, $T_c^{\rm{onset}}$ is $4.75(5)$~K. Here again, the superconducting transition appears relatively broad ($\Delta T_c\approx$~1 K). The in-phase component of the ac susceptibility $\chi'_{\rm{ac}}$ starts to flatten when the resistive transition is complete. The transition is accompanied by a small dissipative peak in the out-of-phase signal $\chi''_{\rm{ac}}$. At the lowest $T$, the diamagnetic screening reaches a value of $98\%$ of the ideal value. 

The temperature dependence of the upper critical field $H_{c2}$ obtained from the resistivity measurements is plotted in Fig.~\ref{Fig3}. Using the values of $H_{c2}$ for $T_c^{\rm{zero}}$ we evaluate the zero temperature limit of the upper critical field to be $\mu_0H_{c2}(0) = 1.42(3)$~T by fitting our data to the generalized Ginzburg-Landau model: $H_{c2}(T) = H_{c2}(0)\left(\frac{1-t^2}{1+t^2}\right)$. The value of $H_{c2}(0)$ is far below the BCS Pauli paramagnetic limit $(B^{\rm{Pauli}}_{c2}=1.83T_c\sim 6.85$~T). Using the zero-temperature relation $\mu_0H_{c2}(0) = \Phi_0/2\pi\xi^2\left(0\right)$, we estimate the coherence length $\xi_0$ to be $15.2(2)$~nm and then using $B_{c2}(0)/B_{\rm{c1}}(0)=2\kappa^{2}/\ln\kappa$ we estimate the Ginzburg-Landau parameter, $\kappa=56.4(8)$ with the penetration depth, $\lambda(0)=860(40)$~nm~\cite{Poole}. The agreement of the transition temperatures measured by both ac and dc susceptibility as well as resistivity is clear evidence of bulk superconductivity in Sn$_{1-x}$In$_{x}$Te. 

\begin{figure}[tb]
\begin{center}
\includegraphics[width=0.9\columnwidth]{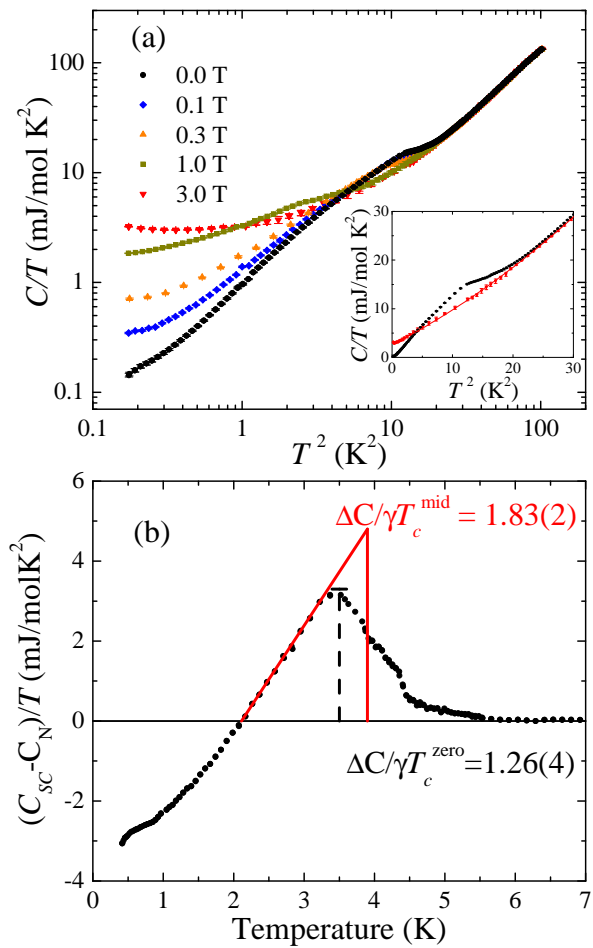}
\caption{\label{Fig4} (Color online). (a) Specific heat divided by temperature ($C/T$) as a function of $T^2$ in various applied fields for superconducting Sn$_{0.6}$In$_{0.4}$Te. The inset shows the region around $T_c$ for the data collected in 0 and 3~T. The solid line shows a fit to $C/T=\gamma+\beta T^2+\delta T^4$ as discussed in the text.  (b) Temperature dependence of the difference between $C$ in the superconducting ($\mu_0H=0$~T) and the normal state ($\mu_0H=3$~T). The peak in the curve occurs a $T_c^{\rm{zero}}=3.5(5)$~K determined from the transport and magnetic susceptibility data. The red line shows an extrapolation of the peak to $T_c^{\rm{mid}}=3.9(5)$~K.}
\end{center}
\end{figure}

Further evidence for the bulk superconductivity is given by the specific-heat $C$. Fig.~\ref{Fig4}a shows $C/T$ against $T^2$ for $x=0.4$ sample in different applied magnetic fields. The inset shows the heat capacity around the transition in zero field and in a field of 3~T, high enough to drive the sample into the normal state. The heat capacity in the normal state can be fit to the expression $C/T=\gamma+\beta T^3+\delta T^4$ where the term linear in $T$ represents the electronic contribution and the higher order terms represent the lattice contribution~\cite{Tari} giving $\gamma=2.62(2)$~mJ/mol K, $\beta=0.635(4)$ mJ/mol K$^4$, $\delta=8.0(1)$~$\mu$J/mol K$^6$. The value of $\beta$ corresponds to a Debye temperature $\Theta_D =183(1)$~K. $\Delta C/\gamma T_c=1.26(4)$ which is slightly smaller than the theoretical Bardeen-Cooper-Schrieffer (BCS) value of 1.43 in the weak-coupling limit. Using the McMillan equation for the superconducting transition temperature $T_c=(\theta_D/1.45)\exp[-1.04\left(1+\lambda_{\rm{e-ph}}\right)/\lambda_{\rm{e-ph}}-\mu^*\left(1+0.62\lambda_{\rm{e-ph}}\right)]$, where the Coulomb pseudo-potential $\mu^*$ is assumed to be 0.13, the electron-phonon coupling constant $\lambda_{\rm{e-ph}}$ is estimated to be 0.67~\cite{McMillan, Poole}. This value of $\lambda_{\rm{e-ph}}$ suggests Sn$_{0.6}$In$_{0.4}$Te should be classified as a weak-intermediate coupling superconductor. A similar disagreement between the strength of the coupling inferred from the jump in heat capacity around $T_c$ and complementary data was noted in studies of superconducting Sn-In-Te doped with much lower levels of In~\cite{Erickson}.

The magnitude of the normal state resistivity of SnTe (with no In) and that of Sn$_{0.6}$In$_{0.4}$Te samples are comparable. The resistance of the brittle Sn$_{0.6}$In$_{0.4}$Te samples increases with thermal cycling, but the form of the $\rho(T)$ curves remains the same. The temperature dependence of the resistivity in SnTe, the parent TCI material, and the resistance ratio $R_{300~\rm{K}}/R_{10~\rm{K}}$ of $\sim17$, indicate that SnTe forms with reasonable number of carriers. The resistance ratio, $R_{300~\rm{K}}/R_{10~\rm{K}}$, of the Sn$_{0.6}$In$_{0.4}$Te sample is only $\sim 1.3$. The resistivity behavior for the $x=0.4$ sample is similar to that observed by Sasaki \textit{et al}.~\cite{Sasaki} for a sample with $x=0.045$ indicating that the behavior remains unchanged when the In substitution levels are increased tenfold. The superconducting behavior seen in the In doped samples (from low levels of around 2\% to levels as high as 40\%, as in the present study) therefore arises out of a poor conductor rather than from a true semiconducting phase. We have also performed resistivity measurements on a range of Sn$_{1-x}$Te$_{1+x}$ samples for $0\leq x\leq0.06$ and found that these variations in the relative Sn/Te content do not result in large changes in the absolute value or temperature dependence of the measured resistivity between 2 and 300~K. 

We briefly address the implications of our results on the search for the occurrence of topological superconductivity in the Sn-In-Te system. Previous studies of the Fermi surface of the $x=0.045$ crystals by Sato \textit{et al}.~\cite{Sato} have provided evidence to show that the band inversion that is observed in the parent TCI material SnTe,  is still present in the Sn-In-Te. The observation of topological surface states in the In substituted SnTe system is promising for the study of topological superconductivity in semiconducting materials with similar characteristics to SnTe. Our results definitely point to the Sn-In-Te materials exhibiting bulk superconductivity and in addition, homogeneous materials can be obtained with ease, in stark contrast to the Cu$_x$Bi$_2$Se$_3$ system. Regarding the pairing mechanisms in play in these superconductors, there are suggestions that the superconductivity in Sn-In-Te may arise from odd-parity pairing ~\cite{Sato, Sasaki}.  LuPtBi, which has been recently proposed to be a topological superconductor with possible odd parity pairing~\cite{Tafti}, crystallizes in a non-centrosymmetric crystal structure and in such systems, generally, mixed singlet-triplet pairing is a possibility. The Sn-In-Te superconductors, however, form in a centro-symmetric crystal structure, and the only evidence for odd parity pairing in the Sn-In-Te comes from the point contact spectroscopy studies by Sasaki \textit{et al}.~\cite{Sasaki}. Our results probe the characteristics of Sn$_{0.6}$In$_{0.4}$Te and confirm that this material is a bulk superconductor and hence a good candidate in the search for 3D topological crystalline superconductivity. Further investigations of the surface states of this superconductor will be required to search for the presence of exotic surface states. The dependence of the superconducting properties on the level of In substitution also needs careful investigation in order to optimize the superconducting characteristics of these materials. These studies are currently underway.

This work was supported by the EPSRC, UK (EP/I007210/1). Some of the equipment used in this research was obtained through the Science City Advanced Materials project: Creating and Characterizing Next Generation Advanced Materials project, with support from Advantage West Midlands (AWM) and part funded by the European Regional Development Fund (ERDF). We wish to thank D. Walker, S. York, and M. Saghir for the x-ray and compositional analysis of the samples used in this study and T. E. Orton for technical support.   

\bibliography{Balakrishnan_v2}\end{document}